\documentclass[a4paper,11pt]{article}
\usepackage{pos}

\usepackage{caption}

\usepackage{graphicx}
\graphicspath{{./figures/}}
\usepackage{subcaption}
\usepackage{dcolumn}

\title{Simulation of the propagation of CR air shower cores in ice}
 \ShortTitle{Simulation of the propagation of CR air shower cores in ice}

\author*[a]{Simon De Kockere}
\author[a]{Krijn de Vries}
\author[a]{Nick van Eijndhoven}

\affiliation[a]{IIHE - Vrije Universiteit Brussel (VUB),\\
  Pleinlaan 2, 1050 Brussels, Belgium}

\emailAdd{simondekockere@gmail.com}

\abstract{Currently new radio detection techniques are being explored to detect astrophysical neutrinos beyond the PeV scale interacting in polar ice. Due to the long attenuation length of radio waves in a medium, it can be expected that such instruments will also be sensitive to the radio emission of cosmic ray air showers. Furthermore, cosmic ray air showers hitting a high-altitude layer of ice will initiate an in-ice particle cascade, also leading to radio emission. We present the first results of detailed simulations of the in-ice continuation of these cosmic-ray-induced particle cascades, using a combination of the CORSIKA Monte Carlo code and the Geant4 simulation toolkit. We give an overview of the general features of such particle cascades and present a parameterization in terms of Xmax of the longitudinal and lateral particle distributions. We discuss the feasibility of observing the in-ice particle cascades, both through the detection of the Askaryan radio emission as well as by using the RADAR reflection technique. Based on these results we find that the expected signals from the continuation of in-ice cosmic-ray induced particle cascades will be very similar to neutrino signals. This means a thorough understanding of these events is necessary in the search for neutrino candidates, while it also promises an interesting in-situ natural calibration source.}

\FullConference{37$^{\rm{th}}$ International Cosmic Ray Conference (ICRC 2021)\\
		July 12th -- 23rd, 2021\\
		Online -- Berlin, Germany}

\begin{document}
\maketitle

\section{Introduction}

A high-energy neutrino interacting in a dense medium like ice will initiate a particle cascade with an excess of negative charges. This negative charge excess will lead to coherent radio emission, as predicted by Askaryan~\cite{Askaryan}. Due to the long attenuation length of radio in ice, this mechanism provides a valuable way of studying astrophysical neutrinos of the highest energies, efficiently extending the energy range of Cherenkov light neutrino observatories like the IceCube Neutrino Observatory. This technique is currently being explored at the Polar regions by several experiments, e.g. \cite{ARA, Arianna, Anita, RNO-G}.

However, also cosmic-ray induced air showers create coherent radio emission reaching the surface, and can thus be expected to form an important background for radio neutrino observatories~\cite{overview_1, overview_2}. Furthermore, at high altitudes of 2 - 3 km, corresponding to the altitudes of \cite{ARA} and \cite{RNO-G}, cosmic-ray induced air showers will typically still have a very energy dense core. The propagation of these energy dense shower cores through the ice will lead to Askaryan radio emission, forming an additional background component.

Only a few studies on the propagation of these shower cores through ice and its corresponding radio emission exist~\cite{Seckel1, Seckel2, Seckel3, cr_askaryan_detectors}, and to fully understand this background component a more detailed analysis is required. Moreover, not only will these shower cores lead to background signals, if well understood they can also serve as a free \emph{in-situ} calibration source. Here we present the first results of a detailed simulation of the propagation of cosmic ray air shower cores in ice and its corresponding radio emission, using a combination of the CORSIKA Monte Carlo code and the Geant4 simulation toolkit. 

\section{Cosmic ray air showers}\label{sec:cosmic_ray_air_showers}


The development of a cosmic ray air shower is a statistical process, resulting in the fluctuation of properties from shower to shower. For illustration, we will discuss the properties of a single proton induced air shower generated with the CORSIKA 7.7100 Monte Carlo code~\cite{corsika}. More precisely, we will discuss its properties around an altitude of 2.4 km, a typical value for the surface of a high-altitude Polar ice sheet. The proton has a primary energy $E_p = 10^{17}$ eV and is generated at a zenith angle $\theta = 0^{\circ}$. We use the QGSJETII-04  high  energy  hadronic  interaction  model, the  GHEISHA  2002d  low  energy  hadronic  interaction model  and  a  MSIS-E-90  atmospheric  model  for  South Pole on December 31, 1997. We apply thinning for electromagnetic particles falling below 10 GeV, using a maximum thinning weight of 10. For hadrons (without $\pi^0$'s) and muons a kinetic energy cut-off of 0.3~GeV was used. For electrons, photons and $\pi^0$'s a kinetic energy cut-off of 0.003~GeV was used.

\begin{figure}
	\centering
	\includegraphics[trim={5.6cm 2cm 5.6cm 2cm},clip,width=0.49\textwidth]{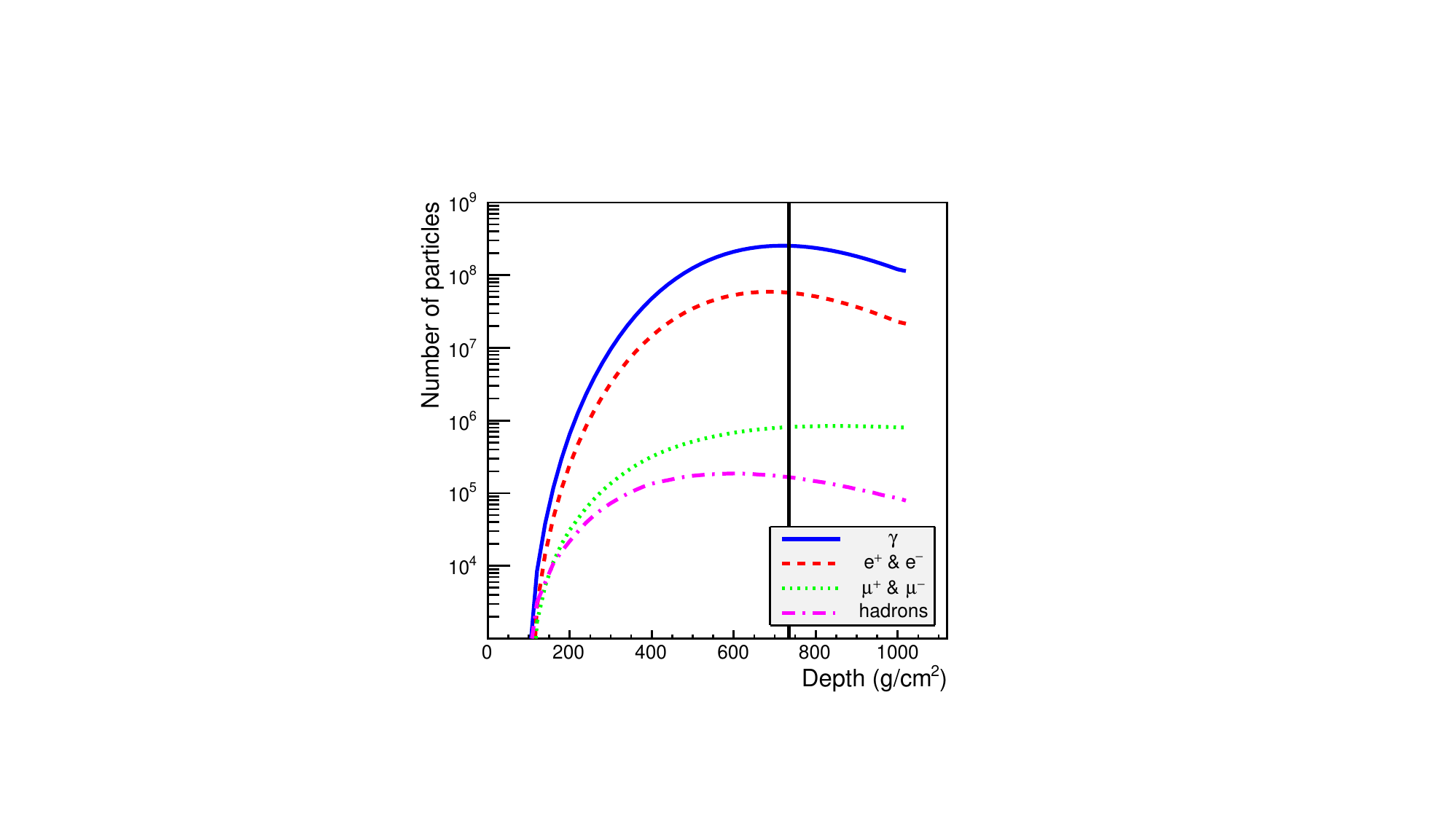}
	\includegraphics[trim={5.6cm 2cm 5.6cm 2cm},clip,width=0.49\textwidth]{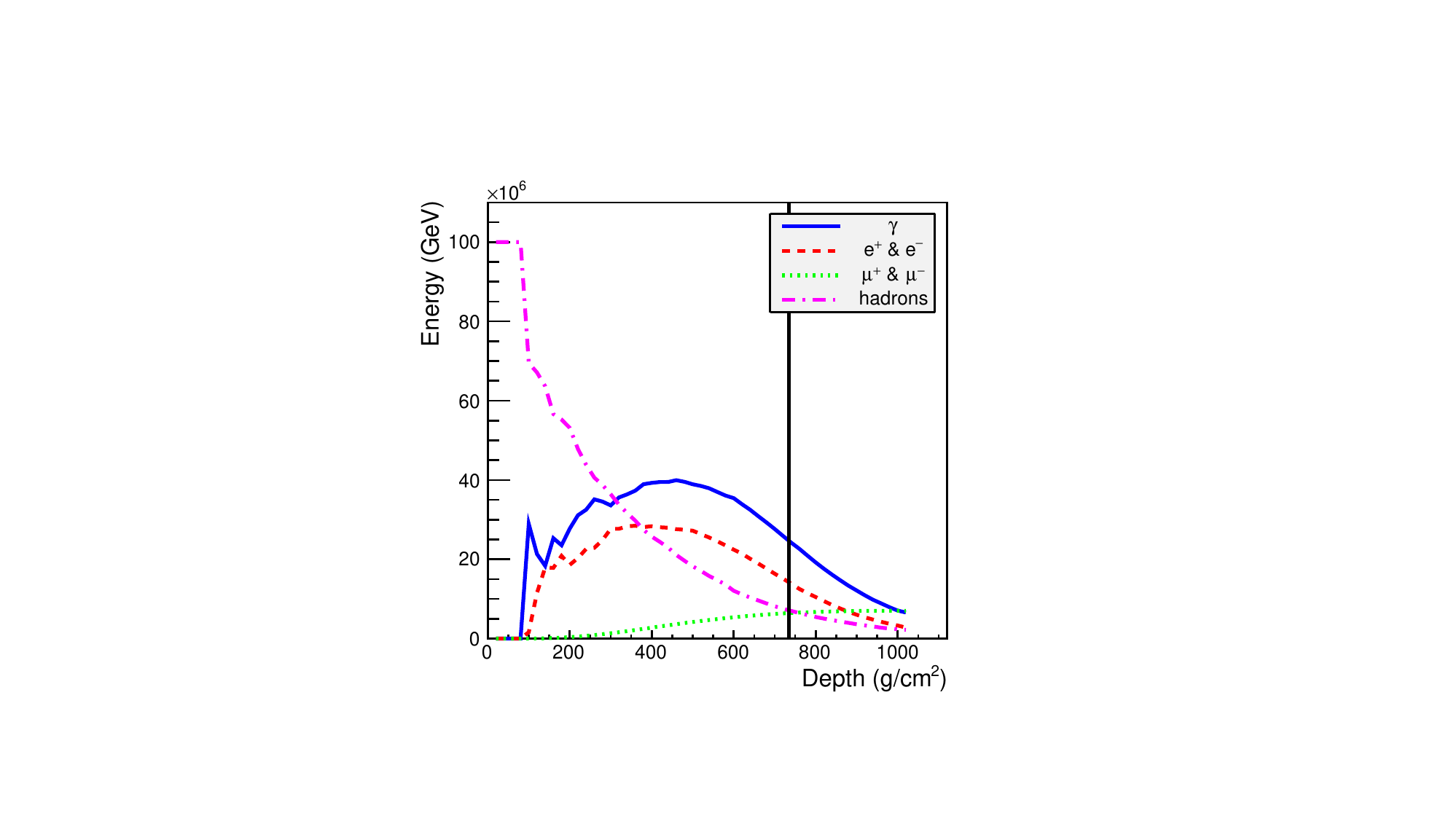}
	\caption{\label{fig:long_profile} The number of particles (\emph{left)} and the distribution of the energy (\emph{right}) of a proton induced air shower, simulated with the CORSIKA Monte Carlo code. The proton has a primary energy $E_p = 10^{17}$~eV and is generated at a zenith angle $\theta = 0^{\circ}$. The shower reaches a maximum number of $e^{+} + e^{-}$ at a depth $X_{max} = 680$ g/cm$^2$. The vertical black line indicates the depth value corresponding to an altitude of 2.4 km.}
\end{figure}

Figure~\ref{fig:long_profile} shows the longitudinal profile of the particle cascade in air. The cascade reaches a maximum number of $e^+ + e^-$ at a depth of $X_{max}$ = 680 g/cm$^2$, which is a good average value, as can be seen in e.g. Figure~2 of~\cite{buitink}. From Figure~\ref{fig:long_profile} it can be seen that the air shower reaches the altitude of 2.4 km right after this shower maximum. At this point in the shower, gamma radiation, electrons and positrons clearly outnumber the muonic and hadronic part. In terms of energy however, the difference is less significant, and the muons and hadrons cannot simply be ignored.

\begin{figure}
	\centering
	\begin{minipage}{0.48\textwidth}
		\centering
		\includegraphics[trim={5.6cm 2cm 5.6cm 2.8cm},clip,width=\textwidth]{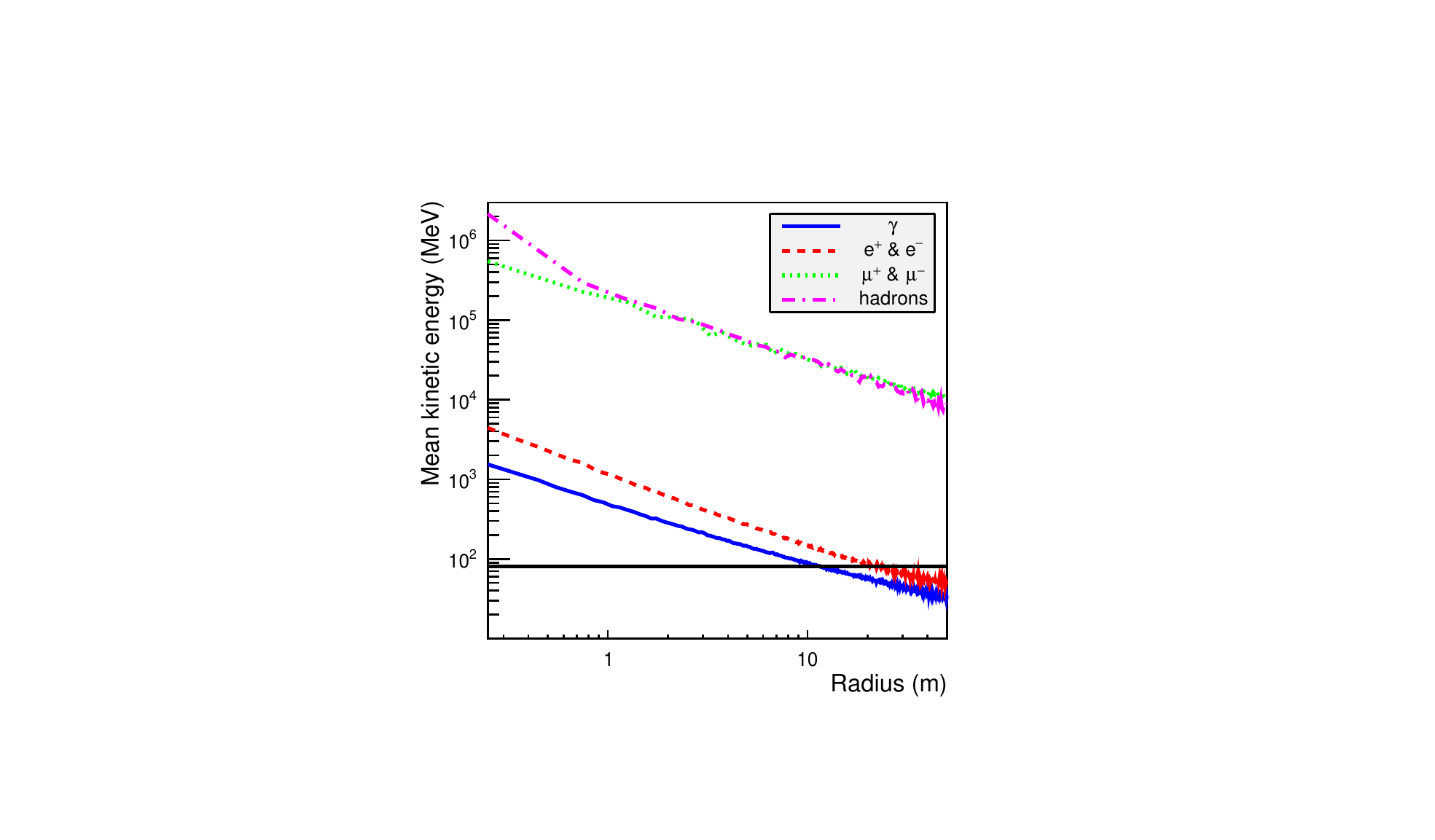}
		\caption{\label{fig:mean_kin_E_dr} The average kinetic energy per particle at an altitude of 2.4 km. For the electromagnetic part ($\gamma$, $e^+$, $e^-$) the average was calculated over radial intervals of $\Delta r = 0.1$ m. For the muonic and hadronic part radial intervals of $\Delta r = 0.5$ m were used. The black horizontal line indicates the critical energy of 80 MeV.}
	\end{minipage}%
	\hspace{1em}
	\begin{minipage}{0.48\textwidth}
		\centering
		\includegraphics[trim={5.6cm 2cm 5.6cm 2.5cm},clip,width=\textwidth]{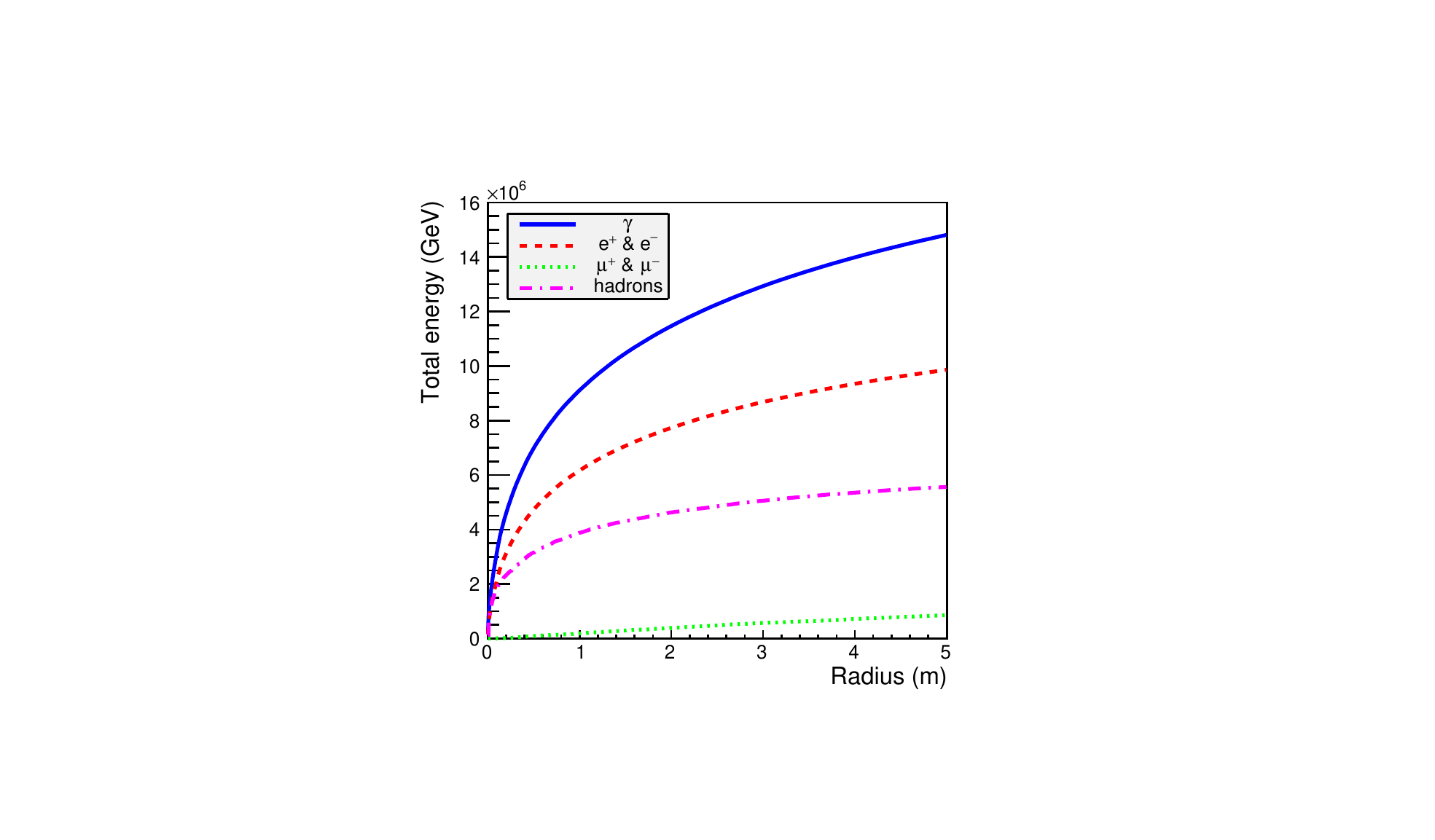}
		\caption{\label{fig:radial_energy_dist} The total energy within a given radius at an altitude of 2.4 km.}
	\end{minipage}
\end{figure}

In Figure~\ref{fig:mean_kin_E_dr} we show the average kinetic energy per particle at an altitude of 2.4 km, calculated over radial intervals of $\Delta r = 0.1$ m. For energies of 80 MeV and above, electromagnetic particles will on average be able to interact, leading to pair-production, continuing the development of the particle cascade. Below this energy ionization losses start to dominate, and the particle cascade dies out. So as indicated in the figure, although the shower reaches this altitude right after shower maximum, we still expect the shower core to develop further, while in total the particle cascade will start to die out. Figure~\ref{fig:radial_energy_dist} shows the total energy within a given radius at the same altitude of 2.4~km. Here we clearly see the shower has a very energy dense core. The first 100~cm of the particle cascade already contains about 15\% of the initial energy of the primary proton.

\section{The in-ice core}\label{sec:in-ice_core}

For the simulation of the propagation of cosmic ray air shower cores in ice, we combine the CORSIKA Monte Carlo code with the Geant4 10.5 simulation toolkit~\cite{geant4}.

Using CORSIKA we calculate the momentum, position and arrival time of all secondary particles reaching an altitude of 2.4~km. This information is passed on to a Geant4 module, which then propagates each secondary particle through ice.

The Geant4 module simulates a volume consisting of multiple horizontal layers of pure ice, each with a constant density and a thickness of 1~cm. The density of the layers follows the density profile measured at the Antarctic Taylor Dome ice cap, given by $\rho(z) = 0.460 + 0.468 \cdot (1 - e^{-0.02z})$, with $\rho$ the density in g/cm$^3$ and $z$ the depth in m. We include the G4EmStandardPhysics constructor, providing us with the standard Geant4 electromagnetic physics, the G4DecayPhysics constructor for the decay physics of long-lived hadrons and leptons and the G4RadioactiveDecayPhysics constructor for radioactive decay physics. We use the default cut-off length of 1~mm for gammas, electrons, positrons and protons, meaning that particles of these types created during the Geant4 simulation are only included in the simulation when they can travel distances larger than this cut-off length. This replaces a hard kinetic energy cut-off.

\begin{figure}
	\centering
	\includegraphics[trim={7.5cm 2cm 3cm 5.7cm},clip,width=0.49\textwidth]{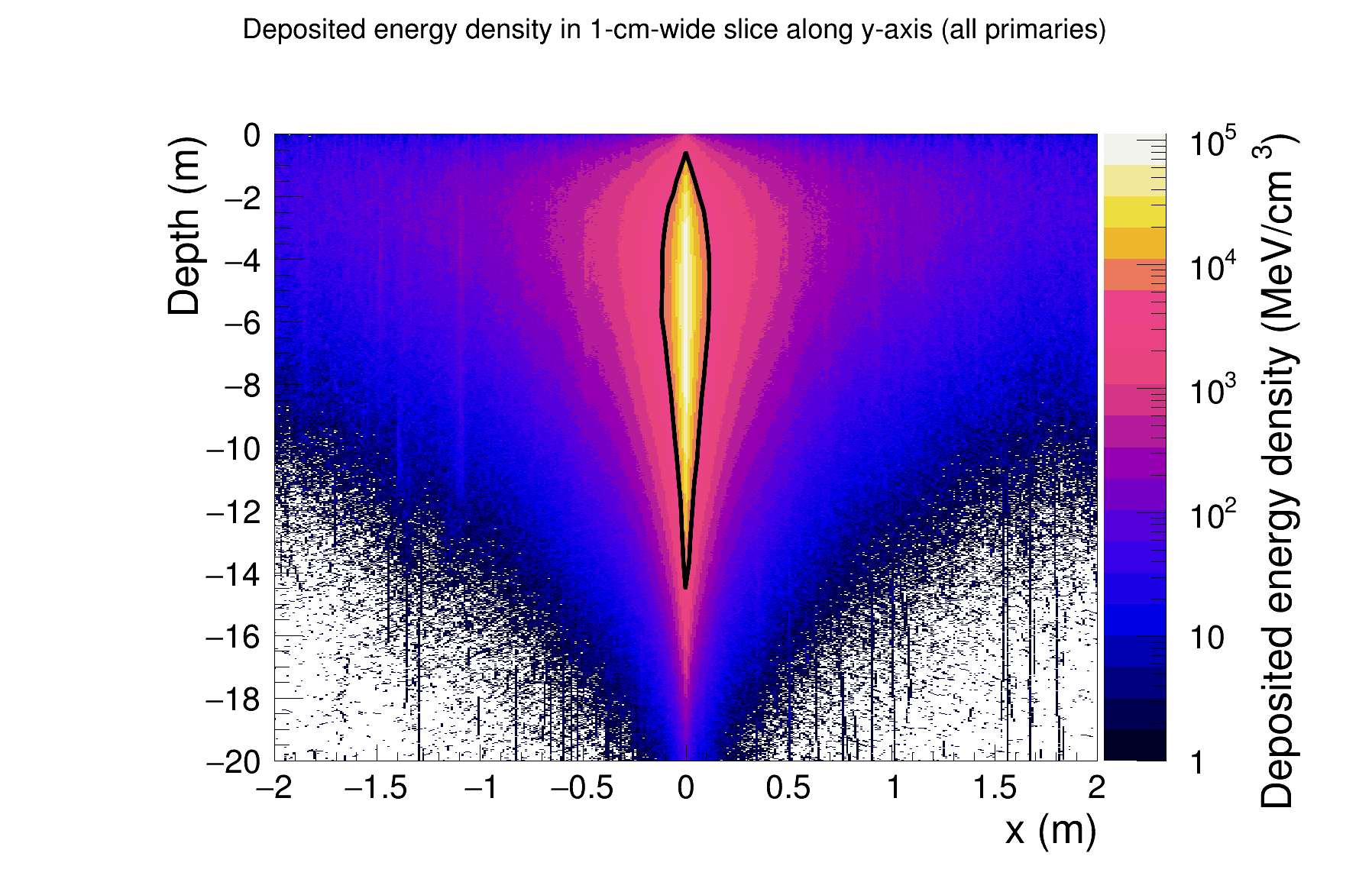}
	\includegraphics[trim={7.5cm 2cm 3cm 5.7cm},clip,width=0.49\textwidth]{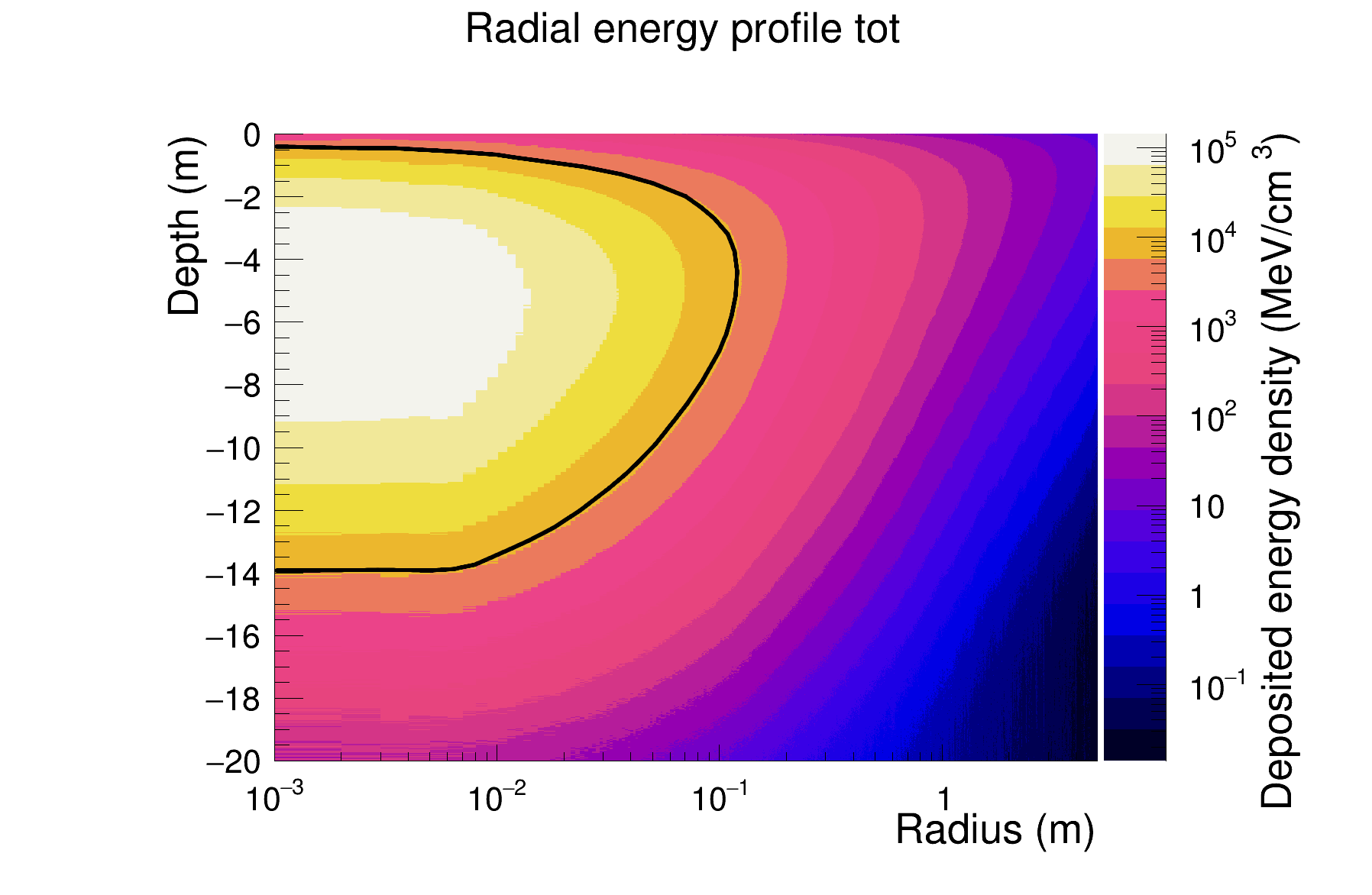}
	\caption{\label{fig:deposited_energy_ice} The deposited energy density in a vertical 1~cm wide slice of the ice going though the center of the particle cascade (\emph{left}) and the deposited energy density in the ice in function of radius and depth (\emph{right}). The black line marks the area for which $\omega_p > 100$~MHz (see Section~\ref{sec:applications}).}
\end{figure}

Figure~\ref{fig:deposited_energy_ice} shows the energy deposited in the ice by the reference shower discussed in Section~\ref{sec:cosmic_ray_air_showers}. We see again that the energy is highly concentrated close to the core of the particle cascade, within a radius of the order of 10~cm, resembling a neutrino induced particle cascade. Furthermore, as discussed above, we can also see that the shower core is still developing when entering the ice.

\begin{figure}
	\centering
	\begin{minipage}{0.48\textwidth}
		\centering
		\includegraphics[trim={5.6cm 2.8cm 5.6cm 2cm},clip,width=\textwidth]{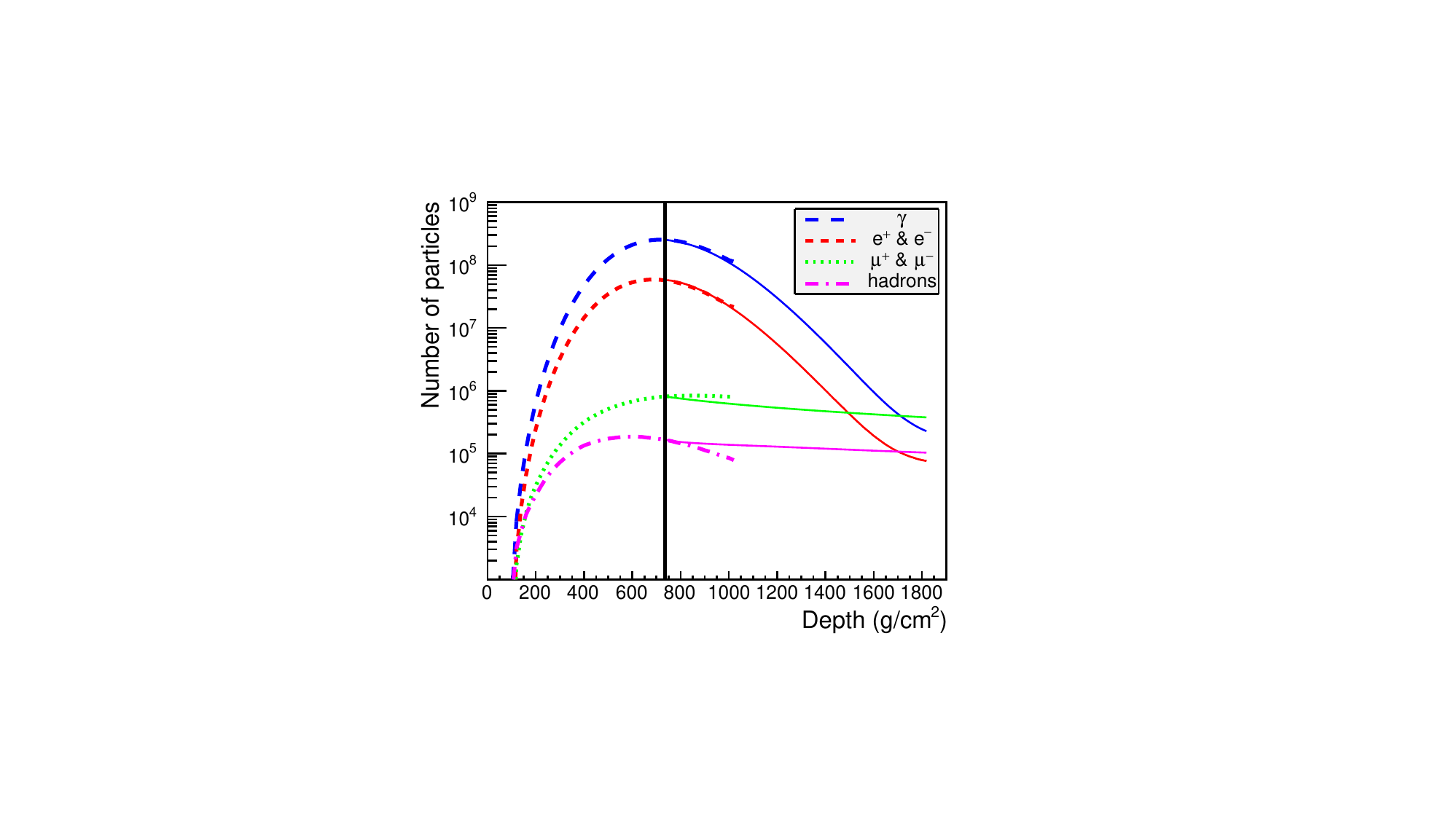}
		\caption{\label{fig:depth_plot} The number of particles in function of depth of the simulated air shower. The dashed lines show the case where the particle cascades propagates through air until reaching sea level. The solid lines show the cascade development when propagated through ice at an altitude of 2.4~km, indicated by the vertical black line. Here we exceptionally use the hard energy cut-offs of CORSIKA in the Geant4 simulation as well.}
	\end{minipage}%
	\hspace{1em}
	\begin{minipage}{0.48\textwidth}
		\centering
		\includegraphics[trim={1.5cm 0.5cm 1cm 1.5cm},clip,width=\textwidth]{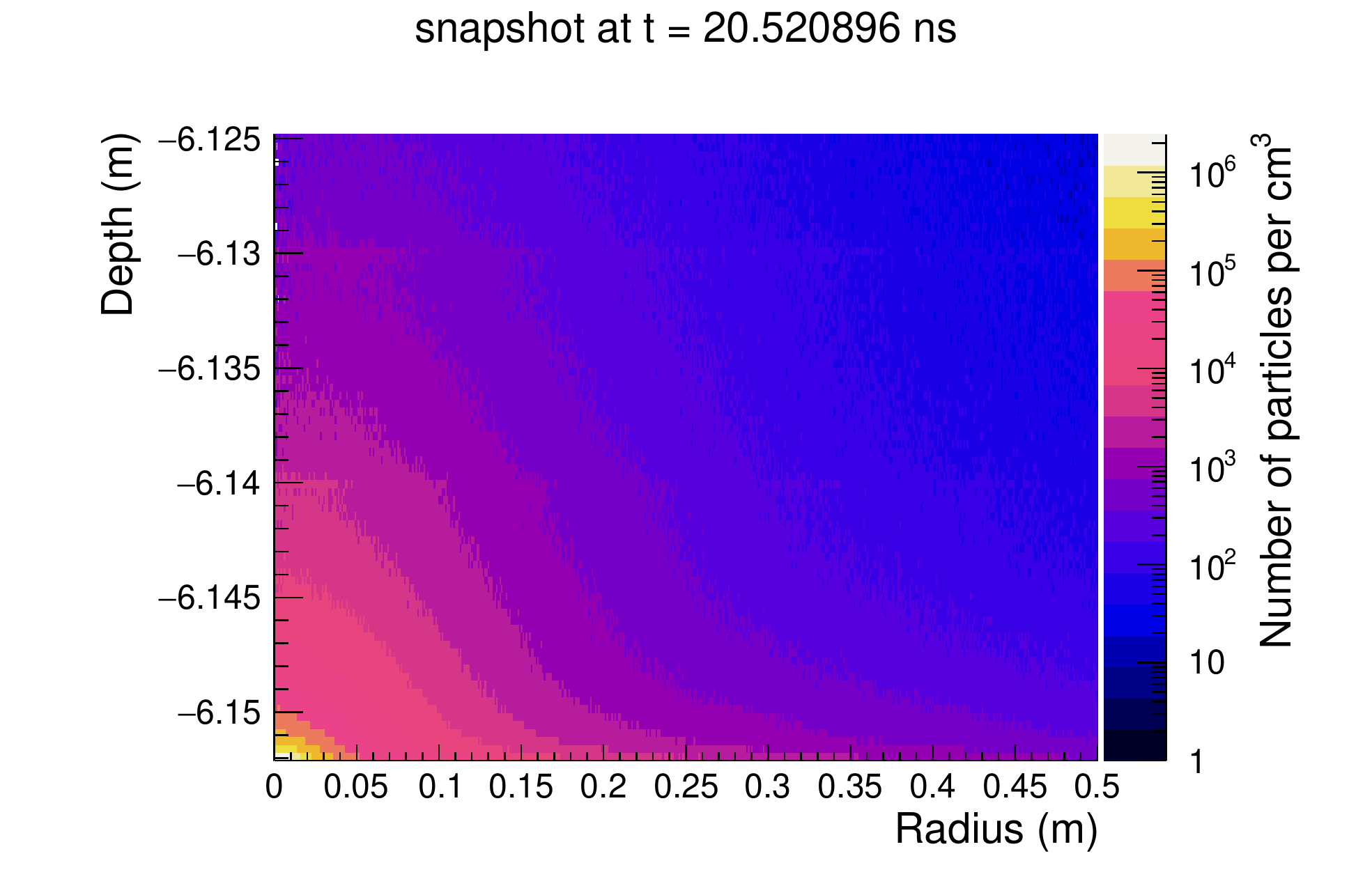}
		\caption{\label{fig:snapshot_proj} The number of particles per cm$^3$ in function of radius and depth at a time $t = 20.5$~ns after the moment of impact of the shower core front on the ice.}
	\end{minipage}
\end{figure}

In Figure~\ref{fig:depth_plot} we show the number of particles in function of depth of the reference shower when propagated through ice, and compare with the case where it continues developing in air instead until reaching sea level. As can be seen, the transition from air into ice does not influence the development of the electromagnetic part of the cascade, and we simply see a continuation of the in-air profile to larger depth values. This means that standard air shower parameterizations such as the Gaisser-Hillas parameterization ~\cite{gaisser-hillas} can be used to describe the electromagnetic part of the cascade in the ice.

Figure~\ref{fig:snapshot_proj} shows a snapshot of the particle cascade at a time $t = 20.5$~ns after the moment of impact of the shower core front on the ice, corresponding with a cascade front depth of 300~g/cm$^2$ with respect to the ice surface. As can be seen, the particles are concentrated in a thin disk, with a thickness of the order of 1~cm. As such, the lateral dimension will be the relevant dimension when studying the corresponding coherent radio emission.

\begin{figure}
	\centering
	\begin{minipage}{0.48\textwidth}
		\centering
		\includegraphics[trim={5.6cm 3cm 5.6cm 2.8cm},clip,width=\textwidth]{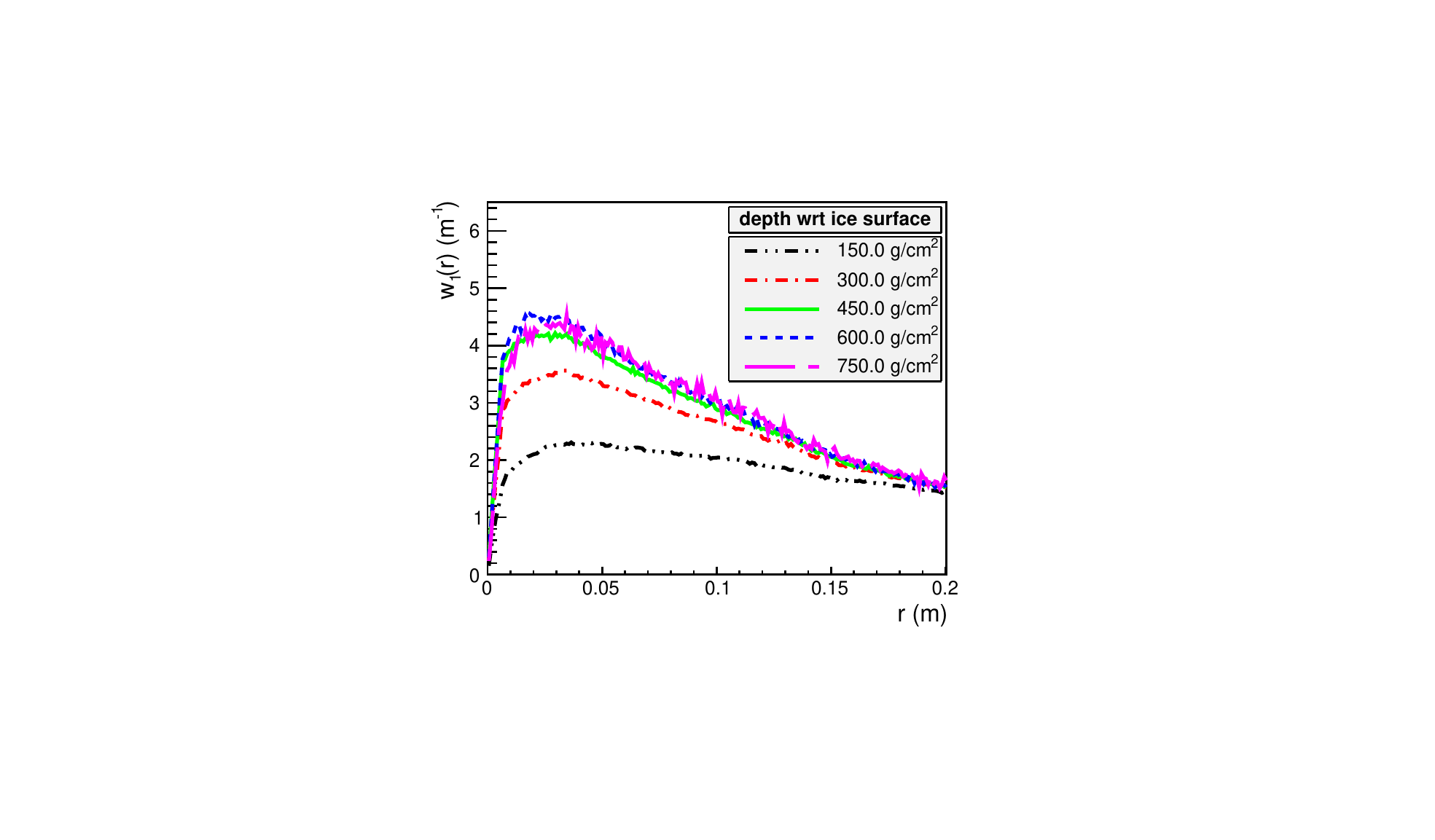}
		\caption{\label{fig:w1} The $w_1(r)$ distributions of the reference shower at different points in time, indicated by the depth of the cascade front with respect to the ice surface.}
	\end{minipage}%
	\hspace{1em}
	\begin{minipage}{0.48\textwidth}
		\centering
		\includegraphics[trim={5.6cm 3cm 5.6cm 2.7cm},clip,width=\textwidth]{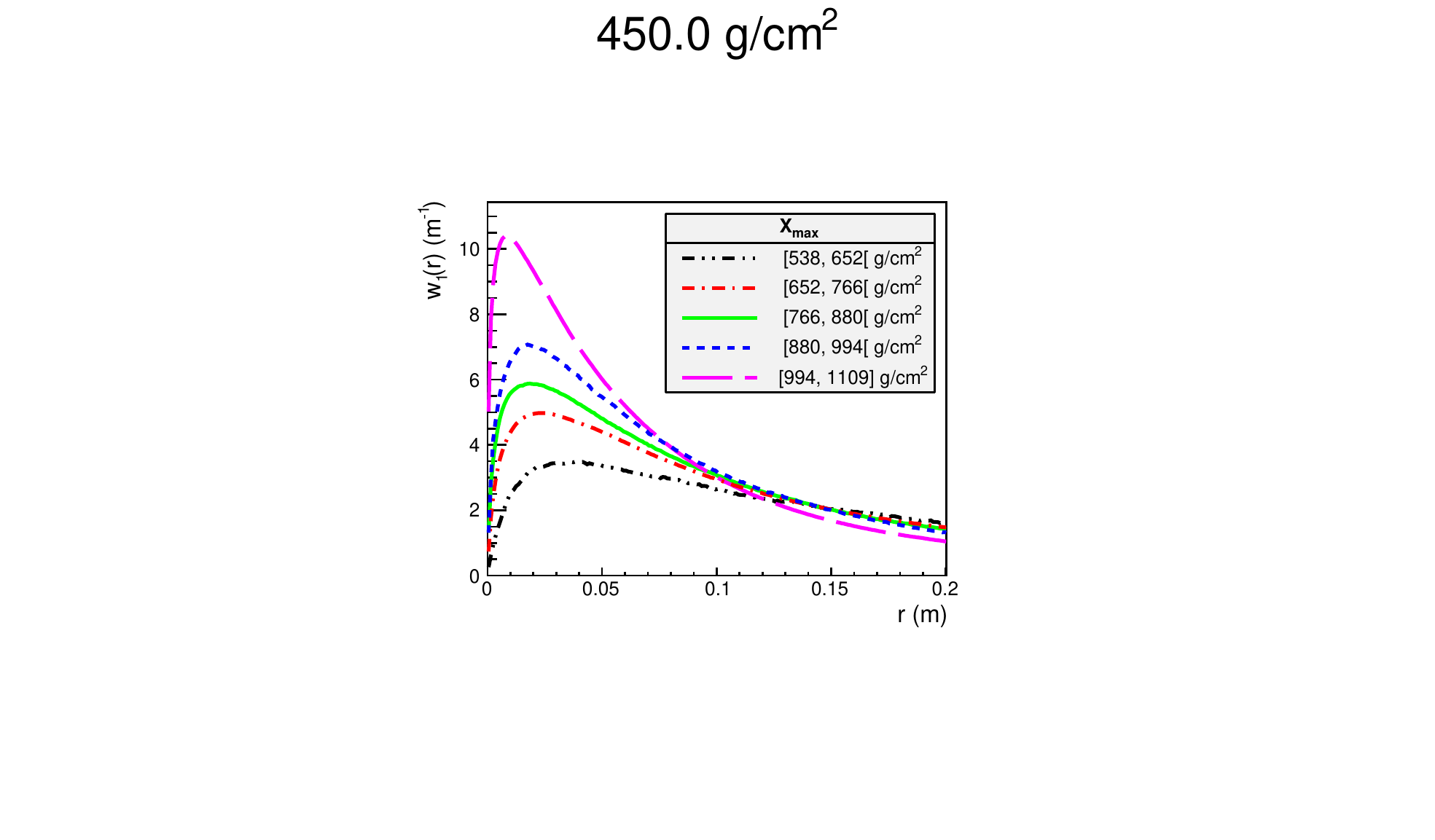}
		\caption{\label{fig:w1_all_showers} The average $w_1(r)$ distributions at a cascade front depth of 450 g/cm$^2$ with respect to the ice surface of the simulation set covering 150 showers. In the legend the lower and upper limits of the different $X_{max}$ groups are given.}
	\end{minipage}
\end{figure}

We describe the lateral particle distribution of the in-ice cascade by constructing the $w_1(r)$ distribution. By definition, $w_1(r)dr$ represents the fractional number of charged particles within the interval $[r, r + dr[$ at a given time, normalized such that the integral from $r = 0$ to $r = 1$ m gives 1. Here, $r$ represents the radius in the shower axis frame. As the cascade develops through time, this distribution will also be time dependent. Figure~\ref{fig:w1} shows the $w_1(r)$ distribution for the reference shower at different points in time, indicated by the depth of the cascade front with respect to the ice surface. They show a peak around $r = 3$~cm, followed by a long tail.

To include a wider range of primary energies and zenith angles, we create 10 different shower sets, each covering primary energies of $10^{16}$~eV up to $10^{18}$~eV in steps of half a decade\footnote{Thinning was applied in CORSIKA for showers with primary energies $E_p$ $\geq$ $10^{17}$ eV on electromagnetic particles falling below $10^{-7}$ $E_p$ with a thinning weight smaller than $10^{-7}$ $E_p[\text{GeV}]$.}, and zenith angles of $0^{\circ}$, $15^{\circ}$ and $30^{\circ}$. This results in 150 showers in total, including the reference shower from Section~\ref{sec:cosmic_ray_air_showers}. Within each set, the random seeds used to run the CORSIKA simulations are identical. In order to derive a simple parameterization of the $w_1(r)$ distributions, we group the showers together based on their $X_{max}$ value, ignoring primary energy and zenith angle. For every group we then construct the $w_1(r)$ distributions for a given cascade front depth, and calculate the mean $w_1(r)$ distribution. The result is shown in Figure~\ref{fig:w1_all_showers}, for a cascade front depth of 450 g/cm$^2$ with respect to the ice surface. As can be seen, the peak value of the $w_1(r)$ distribution correlates with the $X_{max}$ value of the particle shower. A higher $X_{max}$ value implies a denser shower core, resulting in a sharper peak in the $w_1(r)$ distribution.

\begin{figure}
	\centering
	\includegraphics[trim={0 0 0 0},clip,width=\textwidth]{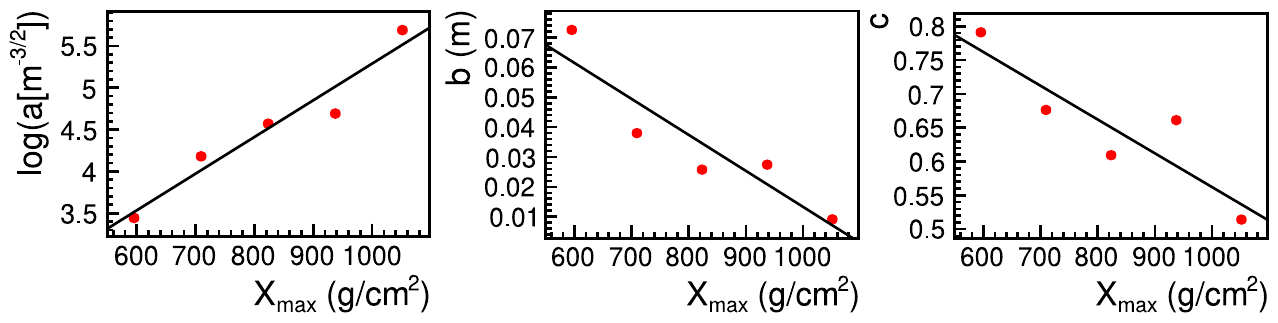}
	\caption{\label{fig:fit_of_fit_parameters} The fit parameters of Equation~\ref{eq:fit_function} for the average $X_{max}$ values of each group, at a cascade front depth of 450 g/cm$^2$ with respect to the ice surface.}
\end{figure}

\begin{figure}
	\centering
	\includegraphics[trim={5.6cm 4cm 5.6cm 2.5cm},clip,width=0.49\textwidth]{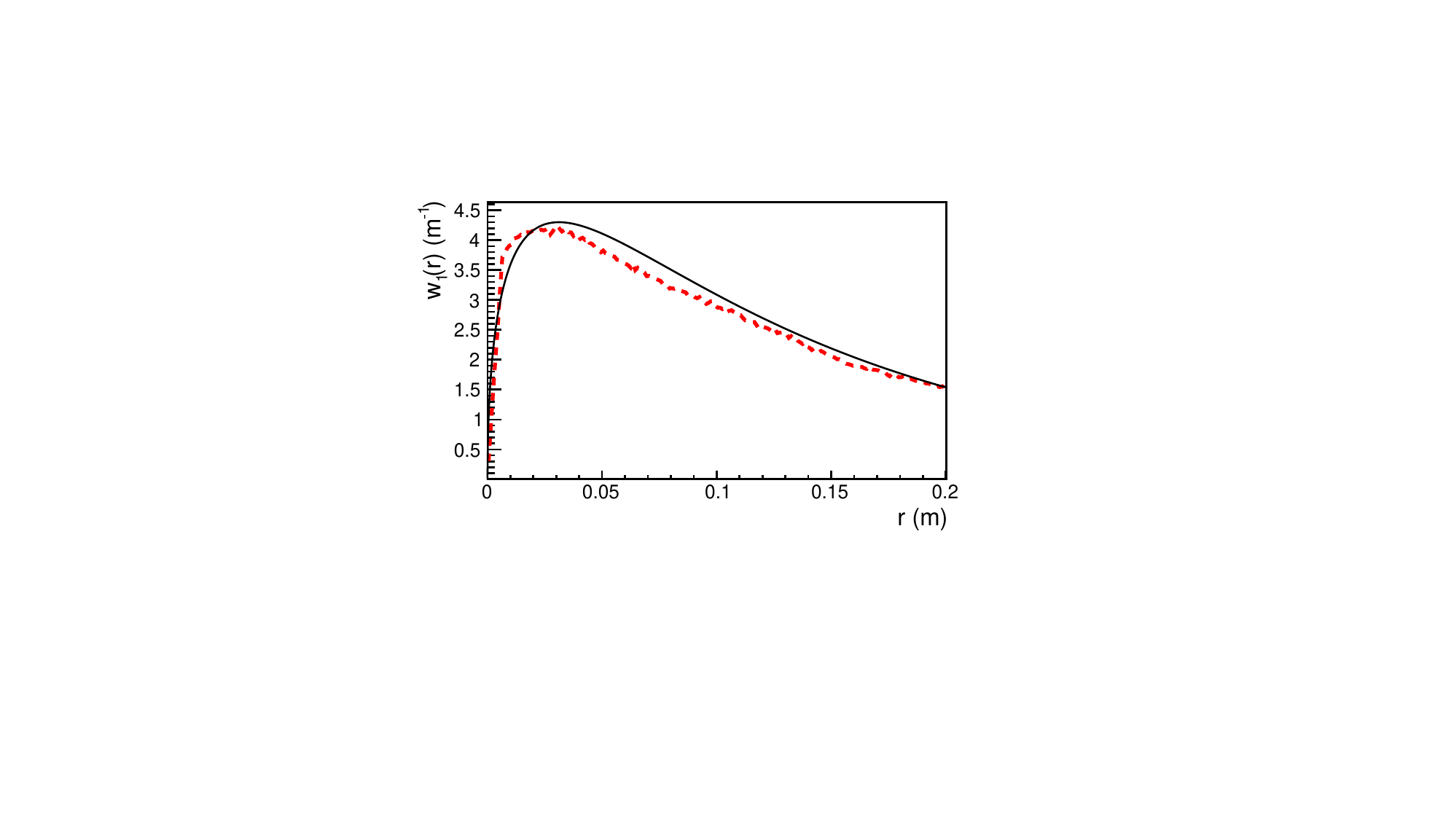}
	\caption{\label{fig:reverse_w1} The $w_1(r)$ distribution at a cascade front depth of 450~g/cm$^2$ with respect to the ice surface for the reference shower from Section~\ref{sec:cosmic_ray_air_showers}. The dashed red line shows the distribution calculated directly from the Geant4 simulation. The solid black line shows the reconstruction of the distribution using Equation~\ref{eq:fit_function} and the linear fits from Figure~\ref{fig:fit_of_fit_parameters}, evaluated at $X_{max} = 680$~g/cm$^2$.}
\end{figure}

We find that the average $w_1(r)$ distributions can be well described by the analytical expression
\begin{equation}\label{eq:fit_function}
	W(r) = a \sqrt{r} e^{-\left(r/b\right)^c},
\end{equation}
with the parameters $a$, $b$ and $c$ depending on the $X_{max}$ value of the shower. Both $\log(a)$, $b$ and $c$ follow a linear relationship with $X_{max}$, as shown in Figure~\ref{fig:fit_of_fit_parameters}. This means that for a shower with known $X_{max}$ value, the parameters $a$, $b$ and $c$ for a given cascade front depth can be constructed, which can be combined with Equation~\ref{eq:fit_function} to get the corresponding $w_1(r)$ distribution. The $w_1(r)$ distribution at a cascade front depth of 450~g/cm$^2$ with respect to the ice surface for the reference shower from Section~\ref{sec:cosmic_ray_air_showers} obtained this way is shown in Figure~\ref{fig:reverse_w1}, where we also compare with the corresponding distribution calculated directly from the Geant4 simulation.

\section{Applications}\label{sec:applications}

To find a first estimate of the Askaryan radio emission associated with the propagation of cosmic ray air shower cores in ice, we implement the end-point formalism~\cite{endpoint} in the Geant4 module, using the code from the work presented in~\cite{anne_zilles} as an example. This formalism is also implemented in the CoREAS extension of CORSIKA, which calculates the radio emission of air showers, and is shown to agree well with experimental results~\cite{coreas}. However, this approach assumes a constant index of refraction $n$ of the medium, which works well for air, but might be an oversimplification for the top layers of natural ice. We use $n = 1.52$.

\begin{figure}
	\centering
	\includegraphics[trim={0 0 0 0},clip,width=\textwidth]{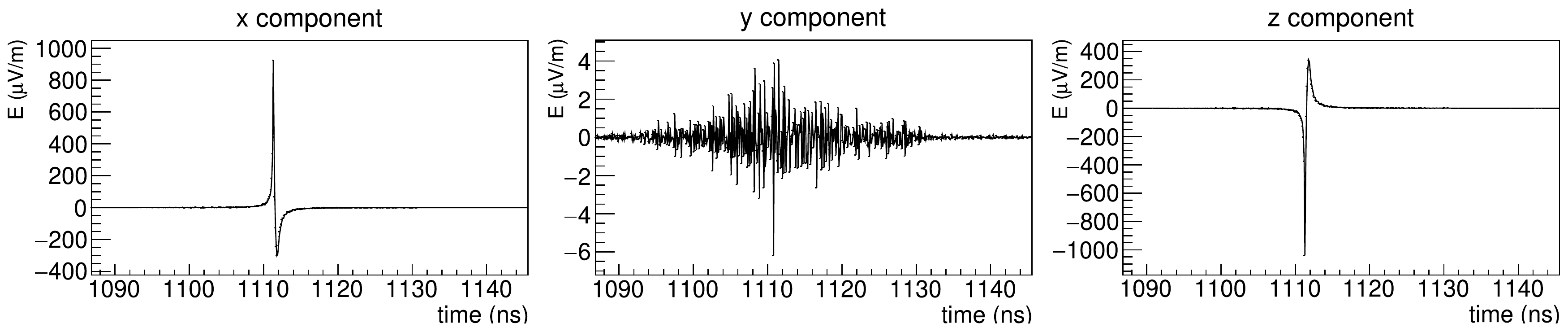}
	\caption{\label{fig:antenna_traces}The electrical field in function of time as seen by a point-like antenna at $x = -160$~m and $z=-150$~m with respect to the point of impact on the ice surface, for the reference shower from Section~\ref{sec:cosmic_ray_air_showers}. It was calculated using the endpoint formalism.}
\end{figure}

Figure~\ref{fig:antenna_traces} shows the electrical field in function of time as seen by a point-like antenna, for the reference shower from Section~\ref{sec:cosmic_ray_air_showers}. The antenna is positioned 150~m deep in the ice ($z = -150$~m), at a horizontal distance of 160~m away from the point of impact of the shower core on the ice surface ($x = -160$~m), which is close to the expected Cherenkov angle. The signal is bipolar and radial polarized, as expected for Askaryan radiation, and reaches a magnitude well above typical detection thresholds of 10-100  $\mu$V/m. Here it should be noted that for a quantitative statement the induced field should be convolved with the antenna, which is foreseen in future work.

As a second application we discuss the feasibility of detecting cosmic ray air shower cores propagating in ice with RADAR reflection techniques~\cite{radar}. The shower core will create a dense plasma in the ice, which should be able to reflect incoming radio waves. We estimate the plasma frequency $\omega_p$ of the plasma created in the ice, indicating the transition between fully coherent and incoherent scattering for a collisionless plasma. Although the in-ice ionization plasma suffers from collisions, as a rule of thumb, signals with a frequency $\omega < \omega_p$ will be reflected by the plasma.

The plasma frequency is given by $\omega_p = 8980 \sqrt{n_q[\text{cm}^{-3}]} \text{ Hz}$, with $n_q$ the free charge density of the plasma. We estimate $n_q$ based on the deposited energy density $\rho_E$ in the ice, using $n_q = \rho_E/(50 \text{ eV})$. As indicated in Figure~\ref{fig:deposited_energy_ice}, the plasma reaches plasma frequencies of 100~MHz and higher, which agrees well with realistic RADAR designs~\cite{ret-cr}.

%
%
%

\end{document}